\documentclass[12pt]{article}
\usepackage{setspace}
\usepackage{amssymb, amsmath, amsthm}
\usepackage[pdftex]{graphicx}
\usepackage{multirow} 
\usepackage{lmodern} 
\usepackage{ifthen} 
\usepackage{rotating} 
\usepackage{mathrsfs}
\usepackage{verbatim}
\usepackage[gen]{eurosym}
\newtheorem{Definition}{Definition}[section]

\newtheorem{Theorem}{Theorem}[section]

\begin{document}

\title{Deciding with Judgment}
\author{Simone Manganelli\footnote{European Central Bank, simone.manganelli@ecb.int. I would like to thank for useful comments and suggestions Andrew Patton, Joel Sobel, Harald Uhlig, as well as participants at the NBER Summer Institute, the Duke/UNC conference on New Developments in Measuring and Forecasting Financial Volatility, the Southampton Finance and Econometrics Workshop, the NBER-NSF Seminar on Bayesian Inference in Econometrics and Statistics in St. Louis, the Berlin conference Modern Econometrics Faces Machine Learning, the $10^{th}$ SoFiE conference in New York, the HeiKaMEtrics Network on Financial Econometrics in Heidelberg, the German Statistical Association in Rostock, as well as seminar participants at the ECB, Humboldt University Berlin, Technische Universitaet Dresden, the University of Bern, the UZH Finance Seminar in Zurich, Tinbergen Institute's Econometrics seminar in Amsterdam, the Finance Seminar Series at the Goethe University in Frankfurt, and the de Finetti Risk Seminars at Bocconi University.}}
\date{March, 2019}
\maketitle

\begin{abstract}
\noindent A decision maker starts from a judgmental decision and moves to the closest boundary of the confidence interval. This statistical decision rule is admissible and does not perform worse than the judgmental decision with a probability equal to the confidence level, which is interpreted as a \emph{coefficient of statistical risk aversion}. The confidence level is related to the decision maker’s aversion to uncertainty and can be elicited with laboratory experiments using urns \`{a} la Ellsberg. The decision rule is applied to a problem of asset allocation for an investor whose judgmental decision is to keep all her wealth in cash.
\end{abstract}



\section{Introduction}
Most people take decisions in an uncertain environment without resorting to formal statistical analysis (Tversky and Kahneman, 1974). I refer to these decisions as \emph{judgmental decisions}. Statistical decision theory uses data to prescribe optimal choices under a set of assumptions (Wald, 1950), but has no explicit role for judgmental decisions. This paper is concerned with the following questions: Is a given judgmental decision optimal in the light of empirical evidence? If not, how can it be improved? 

The answer to the first question is obtained by testing whether, for a given loss function, the first derivative evaluated at the judgmental decision is equal to zero. The answer to the second question is derived from the closest boundary of the confidence interval. The decision rule incorporating judgment is admissible and does not perform worse than the judgmental decision with a probabilty equal to the confidence level. The implication is that abandoning a judgmental decision to follow a statistical procedure always carries the risk of choosing an action worse than the original judgmental decision. This may happen with a probability bounded above by the confidence level.

For concreteness, consider an investor who is about to take the judgmental decision $\tilde{a}$, say, to hold all her assets in cash. She asks an econometrician for advice on whether she should invest some of her money in a stock market index. The best prediction of the econometrician depends on an estimated parameter $\hat{\theta}$, which is affected by estimation risk. For a given utility function provided by the investor, the econometrician can construct a loss function $L(\theta,\tilde{a})$, the loss experienced by the investor if the decision $\tilde{a}$ is taken and the true parameter is $\theta$. Suppose the econometrician is able to recover the distribution of the gradient $\nabla_a L(\hat{\theta}, \tilde{a})$ around the true, but unknown $\theta$. It is possible to test whether the investor's decision $\tilde{a}$ is optimal by testing the null hypothesis that $\nabla_a L(\theta, \tilde{a})$ is equal to zero. If the null hypothesis is not rejected, the econometrician cannot recommend any deviation from $\tilde{a}$. If the null hypothesis is rejected, statistical evidence suggests that marginal deviations from $\tilde{a}$ decrease the loss function relative to $L(\theta,\tilde{a})$. 

\begin{figure}
\caption{Statistical Decision Tree}\label{DecisionTree}
\includegraphics[scale=0.5]{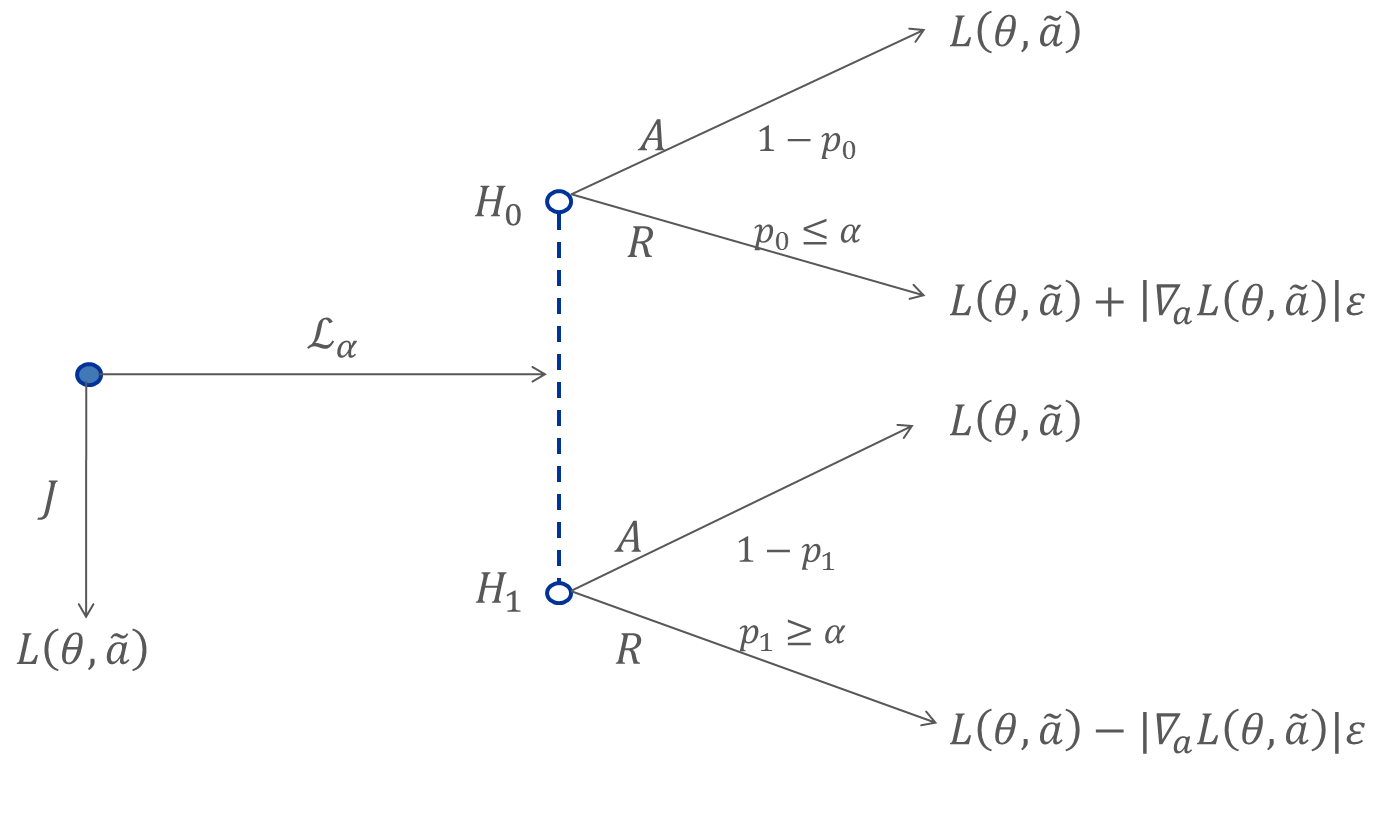}

\footnotesize{\emph{Note}: A decision maker with judgmental decision $\tilde{a}$, confidence level $\alpha$ and loss function $L(\theta,\tilde{a})$ can choose $\tilde{a}$ (branch $J$) or follow a statistical decision rule (branch $\mathcal{L}_{\alpha}$). For a given estimate $\hat{\theta}$ of the statistical parameter, the rule tests whether marginal ($\varepsilon>0$) deviations from $\tilde{a}$ are warranted. It will not decrease the loss if $\tilde{a}$ is optimal (node $H_0$) and it will not increase the loss if $\tilde{a}$ is not optimal (node $H_1$). The dashed line connecting $H_0$ and $H_1$ represents uncertainty, as the decision maker cannot distinguish between the two parts of the tree and no probability can be attached to them. By choosing $\alpha$, she can control the probability $p_0$ of increasing the loss function, in case $H_0$ is true. $\alpha$ provides also the lower bound to the probability $p_1$ of correctly deviating from $\tilde{a}$ in case $H_1$ is true.}
\end{figure}

Denote with $\alpha$ the confidence level used to implement the hypothesis testing. The investor is facing the decision problem depicted in figure \ref{DecisionTree}. The investor has two possible choices. She can hold on to her judgmental decision $\tilde{a}$, denoted by the action $J$, incurring in the loss $L(\theta, \tilde{a})$. Alternatively, she can follow the econometrician's advice, which is equivalent to accepting the bet $\mathcal{L}_{\alpha}$. In this case, she does not know whether she is facing the upper part of the decision tree, denoted by the node $H_0$, or the lower part, denoted by $H_1$. $H_0$ is the unfavorable scenario, in which the null hypothesis is true, so that any deviation from the judgmental decision $\tilde{a}$ results in a higher loss. A marginal $\varepsilon >0$ move away from $\tilde{a}$ results in the loss $L(\theta,\tilde{a}) + |\nabla_aL(\theta,\tilde{a})|\varepsilon$. $H_1$ is the favorable scenario, as one correctly rejects the null hypothesis that $\tilde{a}$ is optimal, producing decisions with lower loss. In this case, a marginal $\varepsilon$ move away from $\tilde{a}$ results in the loss $L(\theta,\tilde{a}) - |\nabla_aL(\theta,\tilde{a})|\varepsilon$. The dash line connecting the two nodes represents true uncertainty for the decision maker, in the sense that it is not possible to attach any probability to being in $H_0$ or in $H_1$. The decision maker can choose the confidence level $\alpha$, which puts an upper bound to the probability that the null is wrongly rejected when it is true. Notice that $\alpha$ represents also the lower bound probability of correctly rejecting $H_0$ when it is false. 

In case of rejection, the investor faces a new, but identical decision problem, except that $\tilde{a}$ is replaced by $\tilde{a} \pm \varepsilon$ (the sign depends on the sign of the empirical gradient). This new action will be rejected if $\nabla_a L(\hat{\theta}, \tilde{a} \pm \varepsilon)$ also falls in the rejection region. Iterating this argument forward, the preferred decision of the investor is the action $\tilde{a}\pm \hat{\Delta}$ which lies at the boundary of the $(1-\alpha)$-confidence interval of $\nabla_a L(\hat{\theta}, \tilde{a} \pm \hat{\Delta})$, the point where the null hypothesis that the decision $\tilde{a}\pm \hat{\Delta}$ is optimal can no longer be rejected. This decision is characterized by the fact that it will produce a loss higher than the original judgmental decision $\tilde{a}$ with probability at most $\alpha$. 

The contribution of this paper lies at the intersection between statistics and decision theory. Statistical decision theory emerged as a discipline in the 1950's with the works of Wald (1950) and Savage (1954). Recent contributions in decision theory focus on modeling behavior when beliefs cannot be quantified by a unique Bayesian prior (Gilboa and Marinacci, 2013) and on models of heuristics describing how people arrive at judgmental decisions (Gennaioli and Shleifer, 2010). This paper, however, is not concerned with the axiomatic foundations of decision theory, but rather with how data can be used to help decision makers improve their judgmental decisions. It falls within Clive Granger's tradition that \emph{`to obtain any kind of best value for a point forecast, one requires a criterion against which various alternatives can be judged'} (Granger and Newbold, 1986, p. 121; see also Granger and Machina, 2006). Recent contributions within this tradition are Patton and Timmermann (2012) and Elliott and Timmermann (2016). Other contributions include Chamberlain (2000) and Geweke and Whiteman (2006), who deal with forecasting using Bayesian statistical decision theory, and Manski (2013 and the references therein), who uses statistical decision theory in the presence of ambiguity for partial identification of treatment response.

The paper is structured as follows. Section 2 sets up the decision environment and introduces the concept of judgment in frequentist statistics. Judgment is defined as a pair formed by a judgmental decision $\tilde{a}$ and a confidence level $\alpha$. Judgment is used to set up the hypothesis to test whether the action $\tilde{a}$ is optimal. Two key results of this section are that the decision rule incorporating judgment is admissible, and that it is either the judgmental decision itself or is at the boundary of the confidence interval of the sample gradient of the loss function.

Section 3 discusses the choice of the confidence level $\alpha$. As illustrated in figure \ref{DecisionTree}, the confidence level $\alpha$ puts an upper bound to the probability that the statistical decision rule performs worse than the judgmental decision. The confidence level can therefore be interpreted as the willingness of the decision maker to take statistical risk and is referred to as the \emph{coefficient of statistical risk aversion}. This concept is closely linked to the idea of ambiguity aversion. The section also discusses how the confidence level $\alpha$ can be elicited with a simple experiment involving urns \`{a} la Ellsberg. 

Section 4 uses an asset allocation problem as a working example to illustrate the empirical performance of various decision rules. Section 5 concludes.

\section{Statistical Decision Rules with Judgment} \label{Decision with judgment}
This section introduces the concept of judgment and shows how hypothesis testing can be used to arrive at optimal decisions. For concreteness, I solve a simple asset allocation problem, but the example can be easily generalized.

Consider an investor holding cash, yielding zero nominal returns. The objective is to minimize a loss function, by deciding what fraction $a \in \mathbb{R}$ to invest in a stock market index, yielding the uncertain return $X$. The decision environment is formally defined as follows.
\begin{Definition}[\textbf{Decision Environment}]\label{Decision Environment}
Let  $\Phi(x)$ denote the cdf of the standard normal distribution. The decision environment is defined by:
\begin{enumerate}
\item $X - \theta \sim \Phi(x)$, where $\theta \in \mathbb{R}$ is unknown.
\item One sample realization $x \in \mathbb{R}$ is observed.\footnote{I denote random variables with upper case letters ($X$) and their realization with lower case letters ($x$).}
\item $a \in \mathbb{R}$ denotes the action of the decision maker.
\item The decision maker minimizes the loss function $L(\theta,a) = -a\theta + 0.5 a^2$.
\end{enumerate}
\end{Definition}

\bigskip
\noindent \textbf{Remark: General case --- } This decision environment can be generalized to cover any continuously differentiable and strictly convex loss function, at the cost of more cumbersome notation. The intuition is the following. Since the main object of interest is the first derivative of the loss function evaluated at $a$ and at the maximum likelihood estimator $\hat{\theta}$, an approximation of the first order conditions around the population parameter $\theta$ gives:
\begin{equation*}
\nabla_{a}L(\hat{\theta}, a) \approx \nabla_{a}L(\theta, a) + \nabla_{a\theta}L(\hat{\theta}, a)(\hat{\theta}-\theta)
\end{equation*}
The statistical properties of the gradient can therefore be deduced from the statistical properties of $\hat{\theta}$. The strict convexity of the loss function guarantees that there is a one to one mapping between $a$ and the gradient (although not linear as in the decision environment above). $\Box$

\bigskip
Consider the following standard definition of a decision rule (Wald, 1950):

\begin{Definition} [\textbf{Decision Rule}]
$\delta(X): \mathbb{R} \rightarrow \mathbb{R}$ is a decision rule, such that if $X=x$ is the sample realization, $\delta(x)$ is the action that will be taken.
\end{Definition}

Classical statistics as developed by Neyman and Fisher has no explicit role for \emph{epistemic} uncertainty (as defined by Marinacci, 2015), as it was motivated by the desire for objectivity. Non sample information is, nevertheless, implicitly introduced in various forms, in particular in the choice of the confidence level and the choice of the hypothesis to be tested.

\subsection{Judgment}
I introduce the following definition of judgment.

\begin{Definition} [\textbf{Judgment}]
\textbf{Judgment} is the pair $A \equiv \{\tilde{a}, \alpha\}$. $\tilde{a} \in \mathbb{R}$ is the \textbf{judgmental decision}. $\alpha \in [0,1]$ is the \textbf{confidence level}.
\end{Definition}

\noindent Judgment is routinely used in hypothesis testing, for instance when testing whether a regression coefficient is statistically different from zero (with zero in this case playing the role of the judgmental decision), for a given confidence level (usually 1\%, 5\% or 10\%). I say nothing about how the judgmental decision is formed. This question is explored by Tversky and Kahneman (1974) and subsequent research. The choice of the confidence level is discussed in section \ref{section_confidenceLevel}. For the purpose of this paper, judgment is a primitive to the decision problem, like the loss function. 
\subsection{Hypothesis Testing}
The decision maker can test whether $\tilde{a}$ is optimal by testing if the gradient $\nabla_a L(\theta,\tilde{a}) = -\theta + \tilde{a}$ is equal to zero. A test statistic for the gradient can be obtained by replacing $\theta$ with its maximum likelihood estimator $X$. 

The novel insight of this paper stems from the realization that the hypothesis to be tested should be conditional on the sample realization $x$. Having observed a negative, say, sample gradient $\nabla_a L(x,\tilde{a})$, one can conclude that values of $a$ higher than $\tilde{a}$ decrease the empirical loss function. The decision maker is interested, however, in the population value of the loss function. If the population gradient is positive, higher values of $a$ would increase the loss function, rather than decrease it. Analogous, but opposite considerations hold if the sample gradient is positive. The null hypothesis to be tested is therefore that the population gradient has opposite sign relative to the sample gradient. For a discussion of the importance of conditioning in statistics, see chapter 10 of Lehmann and Romano (2005) or section 1.6.3 of Berger (1985) and the references therein.

To formalize, partition the sample space according to the sign taken by the sample gradient as follows:
\begin{align}
C_{-} \equiv \{x \in \mathbb{R}: -x+\tilde{a} \le 0\} \\
C_{+} \equiv \{x \in \mathbb{R}: -x+\tilde{a} > 0\} 
\end{align}

\noindent Two cases are possible:

\bigskip
\noindent i) $x \in C_-$, implying that the null hypothesis to be tested is:
\begin{equation} \label{Null1}
H_0: -\theta + \tilde{a} \ge 0 \quad \text{vs} \quad H_1: -\theta + \tilde{a} < 0
\end{equation}

\bigskip
\noindent ii) $x \in C_+$, implying that the null hypothesis to be tested is:
\begin{equation} \label{Null2}
H_0: -\theta + \tilde{a} \le 0 \quad \text{vs} \quad H_1: -\theta + \tilde{a} > 0
\end{equation}
\subsection{Decision}
In an hypothesis testing decision problem, only two actions are possible: The null hypothesis is either accepted or rejected. Let $0\le\gamma\le1$ and $\Phi(c_{\alpha})=\alpha$, and consider again the two cases, conditional on the sample realization $x$. Given the judgment $A=\{\tilde{a},\alpha\},$ define the test functions $\psi_i^A(x), i \in \{-,+\}$ associated with the hypotheses \eqref{Null1}-\eqref{Null2}:

\bigskip
\noindent i) $x \in C_-$ 

\begin{equation}\label{TestFunction1}
  \psi_-^A(x) = \begin{cases}
    0 \quad \text{if }c_{\alpha/2}<-x+\tilde{a}\le0\\
    \gamma \quad \text{if }-x+\tilde{a}=c_{\alpha/2}\\ 
    1  \quad \text{if } -x+\tilde{a}<c_{\alpha/2}
  \end{cases}
\end{equation}

\bigskip
\noindent ii) $x \in C_+$

\begin{equation}\label{TestFunction2}
  \psi_+^A(x) = \begin{cases}
    0 \quad \text{if }0 < -x+\tilde{a} < c_{1-\alpha/2}\\
    \gamma \quad \text{if }-x+\tilde{a}=c_{1-\alpha/2}\\ 
    1  \quad \text{if } -x+\tilde{a}>c_{1-\alpha/2}
  \end{cases}
\end{equation}

\bigskip

The following theorem derives the decision compatible with judgment:
\begin{Theorem}{\textbf{(Decision with judgment)}}\label{Frequentist decision}
Consider the decision environment of Definition \ref{Decision Environment}. A decision maker with judgment $A=\{\tilde{a},\alpha\}$ selects the following decision rule:
\begin{equation}\label{Frequentist decision rule}
\delta^A(X) = I(-X+\tilde{a}\le0) \delta_-^A(X) + I(-X+\tilde{a}> 0) \delta_+^A(X)
\end{equation}
where $I(\cdot)$ is an indicator function which takes value 1 if its argument is true and 0 otherwise,
\begin{align*}
\delta_-^A(X) & = \tilde{a}(1-\psi_-^A(X))+(x+c_{\alpha/2})\psi_-^A(X),  \\
\delta_+^A(X) & = \tilde{a}(1-\psi_+^A(X))+(x+c_{1-\alpha/2})\psi_+^A(X),
\end{align*}
\noindent $c_{\alpha}\equiv \Phi^{-1}(\alpha)$, and $\psi_i^A(X), i\in\{-,+\}$ are the test functions defined in \eqref{TestFunction1}-\eqref{TestFunction2}. 
\end{Theorem}

\noindent \textbf{Proof} --- See Appendix.
\bigskip

The decision rule \eqref{Frequentist decision rule} depends not only on the random variable $X$, but also on the sample realization $x$. To understand the intuition, consider the case i) $x \in C_-$ and the associated null hypothesis $H_0: -\theta+\tilde{a} \ge 0$. The null hypothesis is a statement about the population gradient evaluated at the judgmental decision $\tilde{a}$. It says that \emph{marginally} higher values of $\tilde{a}$ do not decrease the loss function. If it is not rejected at the given confidence level $\alpha$, the chosen action must be $\tilde{a}$. Rejection of the null hypothesis, on the other hand, implies accepting the alternative, which states that \emph{marginally} higher values of $\tilde{a}$ decrease the loss function. Denote the new action marginally away from $\tilde{a}$ with $a_{\varepsilon} = \tilde{a} + \varepsilon$, for $\varepsilon > 0$ and sufficiently small. Notice that $a_{\varepsilon} $ is not random and it is possible to test whether it is optimal, by testing again whether additional marginal moves from $a_{\varepsilon} $ increase the loss function. This reasoning holds for all null hypotheses $H_0: -\theta+a \ge 0$ for any $a \in [\tilde{a}, x+c_{\alpha/2})$. The first null hypothesis which is not rejected is $H_0:-\theta+\hat{a}\ge 0$, where $\hat{a} = x+c_{\alpha/2}$. 

The next theorem shows that this decision cannot be improved.
\begin{Theorem}{\textbf{(Admissibility)}}\label{Admissibility}
The decision $\delta^A(X)$ of Theorem \ref{Frequentist decision} is admissible.
\end{Theorem}

\noindent \textbf{Proof} --- See Appendix.

\bigskip

The admissibility result is a direct consequence of Karlin-Rubin theorem applied to the test functions \eqref{TestFunction1}-\eqref{TestFunction2}. It follows from the fact that the randomness of the decision rule \eqref{Frequentist decision rule} stems from the indicator functions determining the sign of the gradient and from the (conditional) test functions $\psi_i^A(X)$, $i\in\{-,+\}$. The actions to be taken in case of rejection ($x+c_{\alpha/2}$ or $x+c_{1-\alpha/2}$) or non rejection ($\tilde{a}$) are not random.

\section{Choosing the Confidence Level} \label{section_confidenceLevel}
The confidence level $\alpha$ determines the willingness of the decision maker to take statistical risk and therefore I equivalently refer to it as the \emph{coefficient of statistical risk aversion}. The intuition follows from the decision tree of figure \ref{DecisionTree}. A decision maker facing a statistical decision problem is about to take the judgmental decision $\tilde{a}$. The econometrician suggests a statistical decision rule, which by its random nature may perform worse than $\tilde{a}$. The choice of $\alpha$ puts an upper bound to the probability that the statistical decision rule may perform worse than $\tilde{a}$. 

This intuition is formalized by the following theorem. 

\begin{Theorem}[\textbf{Economic interpretation of the confidence level}] \label{Interpretation}
Consider the decision environment of Definition \ref{Decision Environment} and assume the decision maker has judgment $A \equiv \{\tilde{a},\alpha\}$. The decision rule $\delta^{A}(X)$ in \eqref{Frequentist decision rule} performs worse than the judgmental decision $\tilde{a}$ with probability not greater than $\alpha$:
\begin{equation}
P_{\theta}(L(\theta, \delta^{A}(X)) > L(\theta, \tilde{a})) \le \alpha
\end{equation}
\end{Theorem}

\noindent \textbf{Proof} --- See Appendix.

\bigskip

An extremely statistical risk averse decision maker chooses $\alpha = 0$. A zero confidence level results in a degenerate confidence interval which coincides with the entire real line. As a consequence, it is never possible to reject the null hypothesis that the judgmental decision $\tilde{a}$ is optimal. At the other extreme, a statistical risk loving decision maker chooses $\alpha=1$. When $\alpha=1$ the confidence interval degenerates into a single point, which coincides with the \emph{maximum likelihood} decision. In this case, the null hypothesis that $\tilde{a}$ is optimal is always rejected and the decision maker is fully exposed to the possibility that the statistical decision rule will perform worse than the judgmental decision. An intermediate case of statistical risk aversion is represented by the \emph{subjective classical} estimator of Manganelli (2009), which sets $\alpha \in (0,1)$.

The degree of statistical risk aversion $\alpha$ can be elicited with an experiment \`{a} la Ellsberg (1961) where the decision maker has to choose among different couples of urns. Accepting the advice of an econometrician is like accepting a bet with Nature where the probabilities of the payoff are only partially specified.

Consider two urns with 100 balls each. Urn 1 contains only white and black balls, Urn 2 contains white and red balls. If the black ball is extracted, the respondent loses \euro{}100. If the red ball is extracted, the respondent wins an amount in euros which gives an increase in utility equivalent to the reduction in utility produced by the loss of \euro{}100. If the white ball is extracted, nothing happens. The respondent can choose among the composition of the urns described in table \ref{Elicitation}. By accepting one of the bets from 1 to 99, she can control the upper bound probability of losing in case balls are drawn from Urn 1. By choosing this upper bound probability, she automatically chooses the lower bound probability of winning in case the ball is drawn from Urn 2. 

\begin{table}\caption{Experiment to elicit the confidence level $\alpha$}\label{Elicitation}
	\begin{center}
		\begin{tabular}{| c || c | c || c | c |}
		\hline \hline
			& \multicolumn{2}{|c|}{Urn 1} & \multicolumn{2}{|c|}{Urn 2} \\
		\hline
		Bet & White & Black & White & Red \\
		\hline
		0 & 100 & 0 & 100 & 0 \\
		1 & $\ge 99$ & $\le 1$ & $\le 99$ & $\ge 1$ \\
		2 & $\ge 98$ & $\le 2$ & $\le 98$ & $\ge 2$ \\
		... & ... & ... & ... & ...\\
		98 & $\ge 2$ & $\le 98$ & $\le 2$ & $\ge 98$ \\
		99 & $\ge 1$ & $\le 99$ & $\le 1$ & $\ge 99$ \\
		100 & 0 & 100 & 0 & 100 \\
		\hline \hline
		\end{tabular}
	\end{center}
	{\footnotesize{\emph{Note}: The decision maker can choose one of the bets from $0$ to $100$. She will face Urn 1 or Urn 2 with unknown probability. If a white ball is extracted, nothing happens. If a black ball is extracted, the decision maker loses \euro{}100. If a red ball is extracted, she wins a utility equivalent euro amount. Urn 1 and Urn 2 correspond, respectively, to the nodes $H_0$ and $H_1$ of the decision tree in figure \ref{DecisionTree}. The decision maker can partially choose the composition of the urns. For instance, by choosing bet 2, she knows that Urn 1 does not contain more than 2 black balls and Urn 2 contains at least 2 red balls.}}
\end{table}

To understand the link with the statistical decision problem, consider again figure \ref{DecisionTree}. Urn 1 corresponds to node $H_0$ in the upper part of the decision tree in figure \ref{DecisionTree}. Urn 2 corresponds to node $H_1$ in the lower part of the decision tree. The worst case scenario is when the null hypothesis is true, as in this case deviations from $\tilde{a}$ increase the loss. However, even in this case, according to the decision rule \eqref{Frequentist decision rule} there is still the possibility that the null hypothesis is not rejected, in which case the chosen action is $\tilde{a}$. The choice of the confidence level $\alpha$ controls the probability of wrongly rejecting the null. When the null hypothesis is true, it is like having the ball extracted from Urn 1, and choosing $\alpha$ is like choosing the maximum number of black balls contained in Urn 1. The favorable scenario is when the conditional null hypothesis is false. In this case, rejection of the null leads to the choice of a better action, in the sense that it produces a lower loss. When the null hypothesis is false, it is like having the ball extracted from Urn 2. The probability of correctly rejecting the null depends on the power of the test, but is in any case greater than the chosen confidence level $\alpha$. Choosing $\alpha$ is like choosing the minimum number of red balls contained in Urn 2. 

In real world situations, one does not know whether the null hypothesis is true or not, which represents genuine uncertainty and is indicated by the dashed line in figure 1. This is like saying to the participants in the experiment that it is unknown from which urn the ball will be extracted. An extremely statistical risk averse player would always choose not to participate to the bet and retain the judgmental decision $\tilde{a}$, a choice corresponding to bet 0 in the table. A statistical risk loving player would choose bet 100. In general, players with higher degrees of statistical risk aversion would choose bets with lower numbers.

\section{An Asset Allocation Example}
This section implements the decision with judgment, solving a standard portfolio allocation problem. 

The empirical implementation of the mean-variance asset allocation model introduced by Markowitz (1952) has puzzled economists for a long time. Despite its theoretical success, it is well-known that plug-in and Bayesian estimators of the portfolio weights produce volatile asset allocations which usually perform poorly out of sample, due to estimation errors (Jobson and Korkie 1981, Brandt 2007). This paper takes a different perspective on this problem. The decision with judgment provides an asset allocation which does not perform worse than any  given judgmental allocation with a probability equal to the confidence level.

To implement the statistical decision rules, I take a monthly series of closing prices for the EuroStoxx50 index, from January 1999 until December 2015. EuroStoxx50 covers the 50 leading Blue-chip stocks for the Eurozone. The data is taken from Bloomberg. The closing prices are converted into period log returns. Table \ref{table:1} reports summary statistics.

\begin{table}[h!]
\centering
\caption{Summary statistics}
\begin{tabular}{||c c c c c c||} 
 \hline
 Obs & Mean & Std. Dev. & Median & Min & Max \\ [0.5ex] 
 \hline
 206 & -0.06\% & 5.57\% & 0.66\% & -20.62\% & 13.70 \\ [0.5ex] 
 \hline
\multicolumn{6}{p{.75\textwidth}}{\footnotesize{\emph{Note}: Summary statistics of the monthly returns of the EuroStoxx50 index from January 1999 to December 2015.}}

\end{tabular}
\label{table:1}
\end{table}

The exercise consists of forecasting the next period optimal investment in the Eurostoxx50 index of a person who holds \euro{}100 cash. I take the first 7 years of data as pre-sample observations, to estimate the optimal investment for January 2006. The estimation window then expands by one observation at a time, the new allocation is estimated, and the whole exercise is repeated until the end of the sample. 

To directly apply the decision with judgment as discussed in section \ref{Decision with judgment}, which assumes the variance to be known, I transform the data as follows. I first divide the return series of each window by the full sample standard deviation, and next multiply them by the square root of the number of observations in the estimation sample. Denoting by $\{\tilde{x}_t\}_{t=1}^{n}$ the original time series of log returns, let $\sigma$ be the full sample standard deviation and $n_1<n$ the size of the first estimation sample. Then, for each $n_1+s$, $s=0,1,2,...,n-n_1-1$, define:
\begin{equation}
\{x_t\}_{t=1}^{n_1+s} \equiv \{\sqrt{(n_1+s)}\tilde{x}_t/\sigma\}_{t=1}^{n_1+s}   \quad \textrm{and} \quad \bar{x}_{n_1+s} \equiv (n_1+s)^{-1}\sum_{t=1}^{n_1+s}x_t
\end{equation}

I \lq{help}\rq{} the estimates by providing the full sample standard deviation, so that the only parameter to be estimated is the mean return. Under the assumption that the full sample standard deviation is the population value, by the central limit theorem $\bar{x}_{n_1+s}$ is normally distributed with variance equal to one and unknown mean. I can therefore implement the decision rule with judgment, using the single observation $\bar{x}_{n_1+s}$ for each period $n_1+s$. 

\begin{figure}
\caption{Evolution of portfolio values} \label{Values}
\includegraphics[scale=0.47]{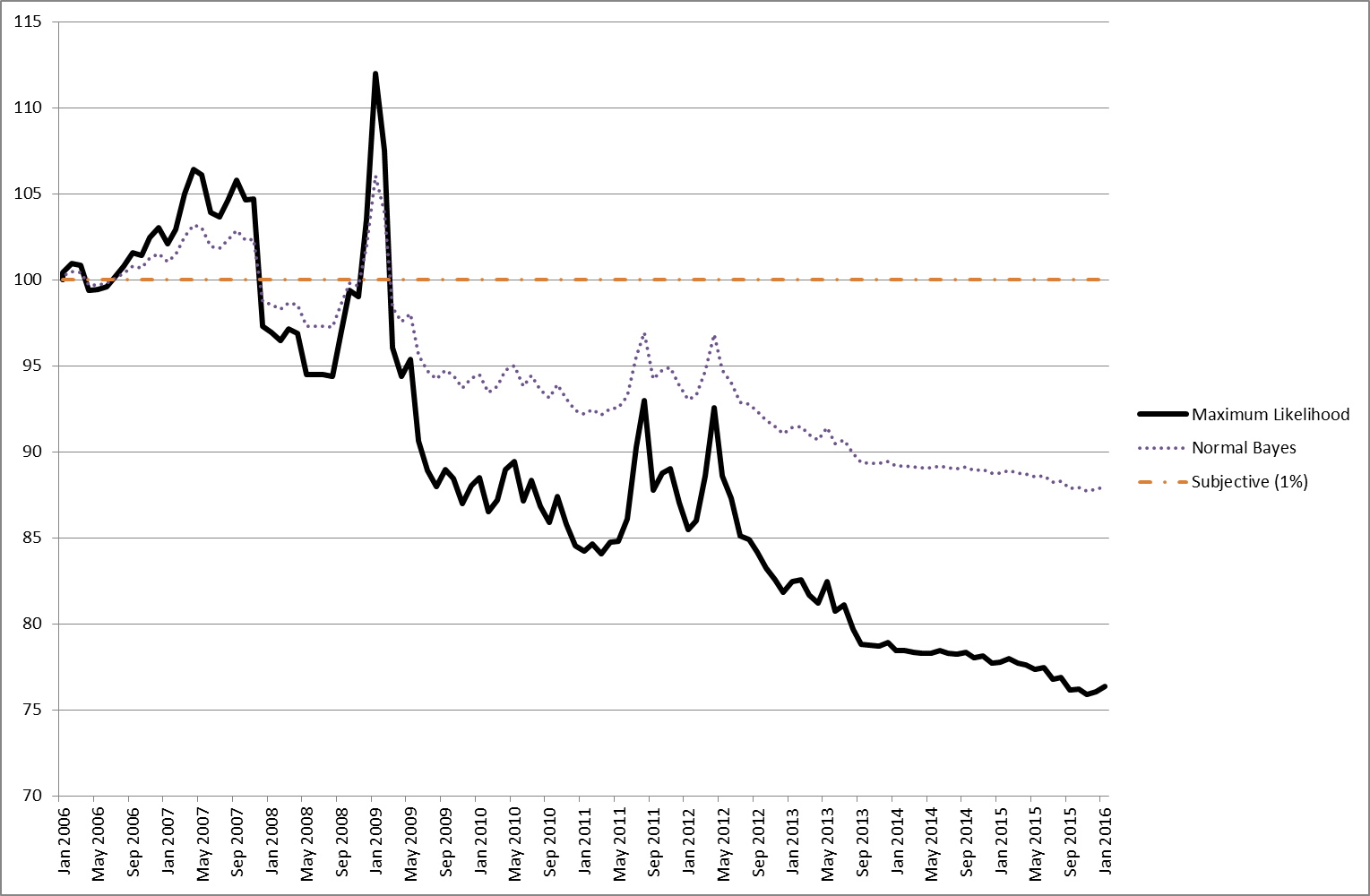}

\footnotesize{\emph{Note:} Time evolution of the value of a portfolio invested in cash and the EuroStoxx50 index following the investment recommendations of the different decision rules.}
\end{figure}

Figure \ref{Values} reports the portfolio values associated with different decision rules. For comparison, I have also included a Bayesian decision with a standard normal prior. Suppose the starting value of the portfolio in January 2006 is \euro{100}. By the end of the sample, after 10 years, an investor using the maximum likelihood decision rule would have lost one quarter of the value of her portfolio. The situation is slightly better with the Bayesian decision rule, as it delivers a loss of around 12\%. The decision with judgment at confidence level at 1\% does not lose anything because it never predicts deviating from the judgmental allocation of holding all the money in cash.

The point of this discussion is not to evaluate whether one decision rule is better than the other, as the decision rules differ only with respect to the choice of the confidence level and the prior, which are both a subjective choice. The purpose is rather to illustrate the implications of choosing different confidence levels. By choosing the maximum likelihood and Bayesian estimators, one has no control on the statistical risk she is going to bear. The decision with judgment, instead, allows the investor to choose a constant probability of underperforming the judgmental allocation: She can be confident that the resulting asset allocation is not worse than the judgmental allocation with the chosen probability. The case of the EuroStoxx50, however, represents only one possible draw, which turned out to be particularly adverse to the maximum likelihood and Bayesian estimators. Had the resulting allocation implied positive returns by the end of the sample, maximum likelihood and Bayesian estimators would have outperformed the decisions with judgment. There is no free lunch: Decision rules with higher statistical risk aversion produce allocations with greater protection to underperformance relative to the judgmental allocation, but also have lower upside potential. In statistical jargon, lower confidence levels protect the decision maker from Type I errors, but imply higher probabilities of Type II errors.

\section{Conclusion}
Judgment plays an important role not just for individuals, but also in policy institutions. Most policy decisions are shaped by the judgment of policy makers. When advising a policy maker, the econometrician can test whether the preferred judgmental decision is supported by data. If not, the decision incorporating judgment is always at the closest boundary of the confidence interval. The probability of obtaining higher losses than those implied by the judgmental decision is bounded by the given confidence level.  

The confidence level reflects the attitude of the decision maker towards statistical uncertainty. I have referred to it as the coefficient of statistical risk aversion. It can be elicited with experiments involving urns \`{a} la Ellsberg. Decision makers characterized by an exteme form of statistical risk aversion (a confidence level equal to 0) always follow their own judgmental decision and ignore the advice of the econometrician. At the other extreme, statistical risk loving decision makers (with a confidence level equal to 1) ignore their judgment and always follow the econometrician's advice, which in this special case corresponds to the maximum likelihood decision. Policy makers are likely characterized by high, but not extreme, coefficients of statistical risk aversion. The framework provided in this paper to measure it may have profound policy implications.

\vspace{6mm}
\noindent \textbf{\Large{Appendix --- Proofs}}
\bigskip 

\noindent \textbf{Proof of Theorem \ref{Frequentist decision} ---} Consider only the case i) $I(-x+\tilde{a} \le 0) =1$. The other case can be proven in a similar way. If $\psi_-^A(x)=0$, the null hypothesis $H_0:-\theta+\tilde{a}\ge0$ is not rejected at the given confidence level $\alpha$. $\tilde{a}$ is therefore retained as the chosen action.

If $\psi_-^A(x)=1$, the null hypothesis is rejected. Rejection of the null implies acceptance of the alternative hypothesis $H_0:-\theta+\tilde{a} \le 0$, that is marginal moves away from $\tilde{a}$ by a sufficiently small amount $\Delta>0$ decrease the loss function. 

Consider now the family of null hypotheses $H_0^{\Delta}: -\theta+\tilde{a} +\Delta \ge 0$ for ${\Delta>0}$. Define also the family of rejection regions $R_{\Delta}\equiv\{y \in \mathbb{R} :-y+\tilde{a}+\Delta< c_{\alpha/2}\}$. Clearly, $x \notin R_{\Delta}$ for any $\Delta \ge \hat{\Delta}\equiv c_{\alpha/2}+x-\tilde{a}$, that is the null hypothesis $H_0^{\Delta}$ is not rejected at the confidence level $\alpha$ for any $\Delta \ge \hat{\Delta}$. 

Denote with $\hat{a}$ the chosen action and suppose that $\hat{a} \ne \tilde{a} + \hat{\Delta}$. If $\hat{a} = \tilde{a} + \Delta < \tilde{a} + \hat{\Delta}$, this implies that $x \in R_{\Delta}$, that is $H_0^{\Delta}: -\theta+\tilde{a} + \Delta\ge0$ is rejected. Therefore this decision cannot be optimal.

If $\hat{a} = \tilde{a} + \Delta > \tilde{a} + \hat{\Delta}$, continuity implies that it exists $\varepsilon>0$ such that the null $H_0^{\Delta-\varepsilon}: -\theta+(\hat{a} + \Delta - \varepsilon) \ge 0$ was rejected at the given confidence level $\alpha$, even though $x \notin R_{\Delta-\varepsilon}$, which implies a contradiction.

The chosen action must therefore be $\hat{a} =\tilde{a} +\hat{\Delta} =c_{\alpha/2}+x$. $\Box$

\bigskip

\noindent \textbf{Proof of Theorem \ref{Admissibility} ---} The risk function of a generic decision $\delta^*$ is:
\begin{align*}
R(\theta, \delta^*) = E_{\theta}(I(-X+\tilde{a} \le 0)R_-(\theta, \delta^*) + I(-X+\tilde{a} > 0) R_+(\theta, \delta^*))
\end{align*}
where 
\begin{align} \label{Risk1}
R_-(\theta, \delta^*) & \equiv E_{\theta|-X+\tilde{a}\le 0}(L(\theta, \delta^*(X))) \\
R_+(\theta, \delta^*) &\equiv E_{\theta|-X+\tilde{a}>0}(L(\theta, \delta^*(X))) \label{Risk2}
\end{align}
and the expectations are taken with respect to the truncated normal distribution.

Consider equation \eqref{Risk1}. The case for equation \eqref{Risk2} can be proven similarly. I prove that the decision $\delta_-^A(X)$ is admissible with respect to the truncated normal distribution. This implies that $R_-(\theta, \delta_-^A) \le R_-(\theta, \delta^*)$ for all $\theta$. Since the same holds for $\delta_+^A(X)$, these two results together imply that $R(\theta,\delta^A) \le R(\theta,\delta^*)$ and therefore that $\delta^A(X)$ is admissible.

To prove that $\delta_-^A(X)$ is admissible with respect to the truncated normal distribution, I verify that the conditions of theorem 4 of Karlin and Rubin (1956) hold. First, note that the truncated normal distribution belongs to the exponential family and therefore possesses a monotone likelihood ratio (see section 1 of Karlin and Rubin, 1956). Second, conditional on observing $-X+\tilde{a} \le 0$, the decision rule of theorem \ref{Frequentist decision} foresees two actions: either the null hypothesis \eqref{Null1} is accepted or rejected. Denote these actions with $a_1$ and $a_2$, respectively. Define the corresponding losses from the original loss function:
\begin{align*}
L_1(\theta) &\equiv  -\tilde{a}\theta + 0.5\tilde{a}^2 \\
L_2(\theta) &\equiv -(x +c_{\alpha/2})\theta + 0.5 (x +c_{\alpha/2})^2
\end{align*}
and note that 
\begin{align*}
L_1(\theta)-L_2(\theta) = -\theta(\tilde{a}-x-c_{\alpha/2})+0.5(\tilde{a}^2-(x+c_{\alpha/2})^2)
\end{align*}
This function is linear in $\theta$ and therefore changes sign only once as a function of $\theta$, specifically at the finite value $\theta=0.5(\tilde{a}+x+c_{\alpha/2})$. Since $\psi_-^A(x)$ is a monotone procedure, the conditions of theorem 4 of Karlin and Rubin (1956) are satisfied and $\delta_-^A(X)$ is admissible. $\Box$

\bigskip

\noindent \textbf{Proof of Theorem \ref{Interpretation}} --- Partitioning the sample space with respect to the gradient $-X+\tilde{a}$:
\begin{align*}
I(L(\theta, \delta^{A}(X)) > L(\theta,\tilde{a})) &= I(-X+\tilde{a} \le 0) I(L(\theta, \delta^{A}_-(X)) > L(\theta,\tilde{a})) \\
	&+I(-X+\tilde{a} > 0) I(L(\theta, \delta^{A}_+(X)) > L(\theta,\tilde{a}))
\end{align*}
\noindent Consider again only the case i) $I(-X+\tilde{a} \le 0) = 1$, as the other one is similar. 

Let us find out first the values of $a$ for which $L(\theta, a) > L(\theta,\tilde{a})$. This is equivalent to finding out when the function $-a\theta + 0.5a^2 + \tilde{a}\theta - 0.5\tilde{a}^2$ is positive, which it is for $a<\theta - |-\theta + \tilde{a}|$ and $a>\theta + |-\theta + \tilde{a}|$. Therefore: 
\begin{align}\nonumber 
I&(-X+\tilde{a} \le 0) I(L(\theta, \delta^{A}_-(X)) > L(\theta,\tilde{a})) = \\
&= I(-X+\tilde{a} \le 0) (I(\delta^{A}_-(X)<\theta - |-\theta + \tilde{a}|) + \label{P1} \\
& \qquad \qquad \qquad \quad + I(\delta^{A}_-(X)>\theta + |-\theta + \tilde{a}|)) \label{P2}
\end{align}
and note also that $\delta^{A}_-(X) = \tilde{a} +(x+c_{\alpha/2}-\tilde{a})\psi_-^A(X)$.

Suppose first that $-\theta + \tilde{a}>0$. Substituting the decision rule and rearranging the terms, the term \eqref{P1} is equal to:
\begin{align*}
I(\delta^{A}_-(X)<\theta - |-\theta + \tilde{a}|)&=I(\delta^{A}_-(X)<2\theta-\tilde{a})\\
&= I((x+c_{\alpha/2}-\tilde{a})\psi_-^A(X) < 2\theta - 2\tilde{a}) \\
&=0
\end{align*}
because $(x+c_{\alpha/2}-\tilde{a})\psi_-^A(X) \ge 0$, and the term \eqref{P2} is equal to:
\begin{align*}
I&(-X+\tilde{a}\le 0)I(\delta^{A}_-(X)>\tilde{a})\\
&\quad =  I(-X+\tilde{a}\le 0)I((x+c_{\alpha/2}-\tilde{a})\psi_-^A(X) > 0) \\
&\quad =  I(-X+\tilde{a}\le 0) I(-X+\tilde{a}<c_{\alpha/2})\\
&\quad =I(-X+\theta<c_{\alpha/2}+\theta-\tilde{a}) \\
&\quad \le I(-X+\theta<c_{\alpha/2})
\end{align*}
where the inequality follows from the fact that the case currently analyzed is $-\theta+\tilde{a}>0$.

If $-\theta + \tilde{a}<0$, similar reasoning can be used to show that the terms \eqref{P1} and \eqref{P2} are equal to:
\begin{align*}
I(\delta^{A}_-(X)<\tilde{a})&= 0 \\
I(-X+\tilde{a}\le 0)I(\delta^{A}_-(X)>2\theta - \tilde{a}) &\le I(-X+\theta<c_{\alpha/2})
\end{align*} 

Combining all these results gives:
\begin{align*}
E_{\theta}(I(L(\theta, \delta^{A}(X)) > L(\theta,\tilde{a}))) &\le E_{\theta} (I(-X+\theta<c_{\alpha/2}) + I(-X+\theta>c_{1-\alpha/2})) \\
	&=\alpha \quad \Box
\end{align*}

\vspace{15mm}

\noindent \textbf{\Large{References}}

\bigskip
\noindent Berger, J. O. (1985), \emph{Statistical Decision Theory and Bayesian Analysis} (2nd ed.), New York: Springer-Verlag.

\bigskip
\noindent Brandt, M.W. (2009), Portfolio Choice Problems, in Y. Ait-Sahalia and L. P. Hansen (eds.), \emph{Handbook of Financial Econometrics}, North Holland.

\bigskip
\noindent Chamberlain, G., (2000), Econometrics and decision theory, \emph{Journal of Econometrics}, 95 (2), 255-283.

\bigskip
\noindent G. Elliott and A. Timmermann (2016), \emph{Economic Forecasting}, Princeton University Press.

\bigskip
\noindent Ellsberg, D. (1961), Risk, Ambiguity, and the Savage Axioms, \emph{The Quarterly Journal of Economics}, 75 (4), 643–669.

\bigskip
\noindent Gennaioli, N. and A. Shleifer (2010), What Comes to Mind, \emph{The Quarterly Journal of Economics}, 125 (4), 1399–1433.

\bigskip
\noindent Gilboa, I. and M. Marinacci (2013), Ambiguity and the Bayesian Paradigm, in \emph{Advances in Economics and Econometrics: Theory and Applications}, Tenth World Congress of the Econometric Society, D. Acemoglu, M. Arellano and E. Dekel (Eds.), New York, Cambridge University Press.

\bigskip
\noindent Geweke, J. and C. Whiteman (2006), Bayesian Forecasting, in \emph{Handbook of Economic Forecasting, Volume I}, edited by G. Elliott, C. W. J. Granger and A. Timmermann, Elsevier.

\bigskip
\noindent Granger, C.W.J. and M.J. Machina (2006), Forecasting and decision theory, in G. Elliott, C. Granger and A. Timmermann (eds.), \emph{Handbook of Economic Forecasting}, vol.1, 81-98, Elsevier.

\bigskip
\noindent Granger, C.W.J. and P. Newbold (1986), \emph{Forecasting Economic Time Series}, Academic Press.

\bigskip
\noindent Jobson, J.D. and B. Korkie (1981), Estimation for Markowitz Efficient Portfolios, \emph{Journal of the American Statistical Association}, 75, 544-554.

\bigskip
\noindent Karlin, S. and H. Rubin (1956), The Theory of Decision Procedures for Distributions with Monotone Likelihood Ratio, \emph{The Annals of Mathematical Statistics}, 27, 272-299.

\bigskip
\noindent Lehmann, E.L.  and J.P. Romano (2005), \emph{Testing Statistical Hypothesis}, Springer.

\bigskip
\noindent Manganelli, S. (2009), Forecasting with Judgment, \emph{Journal of Business and Economic Statistics}, 27 (4), 553-563.

\bigskip
\noindent Manski, C.F. (2013), \emph{Public policy in an uncertain world: analysis and decisions}, Harvard University Press.

\bigskip
\noindent Markowitz, H.M. (1952), Portfolio Selection, \emph{Journal of Finance}, 39, 47-61.

\bigskip
\noindent Marinacci, M. (2015), Model Uncertainty, \emph{Journal of European Economic Association}, 13, 998-1076.

\bigskip
\noindent Savage, L.J. (1954), \emph{The Foundations of Statistics}, New York, John Wiley \& Sons.

\bigskip
\noindent Patton, A.J. and A. Timmermann (2012), Forecast Rationality Tests Based on Multi-Horizon Bounds, \emph{Journal of Business and Economic Statistics}, 30 (1), 1-17.

\bigskip
\noindent Tversky, A. and D. Kahneman (1974), Judgment under Uncertainty: Heuristics and Biases, \emph{Science}, 1124-1131.

\bigskip
\noindent Wald, A. (1950), \emph{Statistical Decision Functions}, New York, John Wiley \& Sons.

\end{document}